\documentclass{JINST}
\usepackage{subfig}

\newcommand{\be}{\begin{equation}}
\newcommand{\ee}{\end{equation}}
\newcommand{\bea}{\begin{eqnarray}}
\newcommand{\eea}{\end{eqnarray}}

\newcommand{\nl}{$\,$\\}

\graphicspath{{figures/}}

\title{APDs as Single-Photon Detectors for Visible and Near-Infrared Wavelengths down to Hz Rates}
\date{September 22, 2011}

\author{R.~J\"ohren$^a$\thanks{Corresponding author.}, R.~Berendes$^a$, W.~Buglak$^a$, D.~Hampf$^a$, V.~Hannen$^a$, J.~Mader$^a$, W.~N\"ortersh\"auser$^{b,c}$, R.~S\'anchez$^c$ and C.~Weinheimer$^a$\\
\llap{$^a$}Institut f\"ur Kernphysik, Westf\"alische Wilhelms-Universit\"at M\"unster,
  48149 M\"unster, Germany\\
\llap{$^b$}Institut f\"ur Kernchemie, Johannes Gutenberg Universit\"at Mainz, 55128 Mainz, Germany\\
\llap{$^c$}GSI Helmholtzzentrum f\"ur Schwerionenforschung, Darmstadt, Germany\\

  E-mail: \email{rjoehren@uni-muenster.de}
}

\abstract{
For the SPECTRAP experiment at GSI, Germany, detectors with single-photon counting capability in the visible and near-infrared regime are required. For the wavelength region up to 1100~nm we investigate the performance of $2\times2$~mm$^2$ avalanche photo diodes (APDs) of type S0223 manufactured by Radiation Monitoring Devices. To minimize thermal noise, the APDs are cooled to approximately -170$^{\circ}$C using liquid nitrogen. By operating the diodes close to the breakdown voltage it is possible to achieve relative gains in excess of $2\cdot10^4$. Custom-made low noise preamplifiers are used to read out the devices. 
The measurements presented in this paper have been obtained at a relative gain of $2.2\cdot10^4$. At a discriminator threshold of 6~mV the resulting dark count rate is in the region of 230~s$^{-1}$. With these settings the studied APDs are able to detect single-photons at 628 nm wavelength with a photo detection efficiency of $(67\pm7)$\%. 
Measurements at 1020~nm wavelength have been performed using the attenuated output of a grating spectrograph with a light bulb as photon source. With this setup the photo detection efficiency at 1020~nm has been determined to be $(13 \pm 3)$\%, again at a threshold of 6~mV.
}


\begin{document}
\section{Introduction}
\label{sec::introduction}
\begin{figure}[h]
\centering
\includegraphics[width=0.5\textwidth]{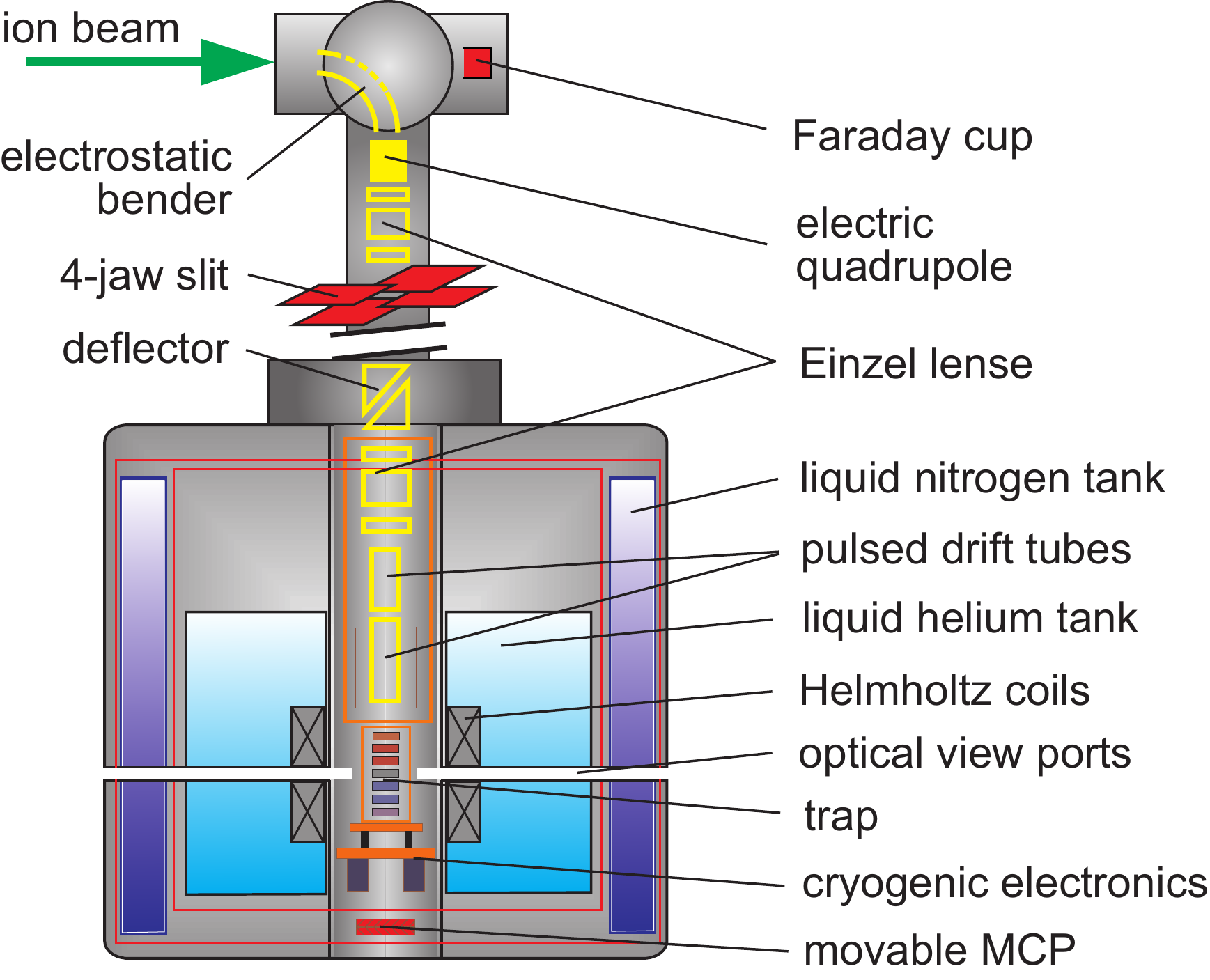}
\caption[SpecTrap setup scheme]{Overview of the SPECTRAP setup (not to scale). Ions are injected either from HITRAP or from an off-line/local ion source and are captured in the central Penning trap located inside two superconducting magnets in a Helmholtz configuration. Light from tunable lasers can enter the trap from below the setup once the multichannel plate (MCP) detector which is used for diagnostic purposes is removed.
Fluorescence photons can be detected through two pairs of optical view ports~\cite{And10}, perpendicular to each other.}
\label{fig::setup}
\end{figure}

\begin{figure}[htbp!]
\centering
\includegraphics[width=0.6\textwidth]{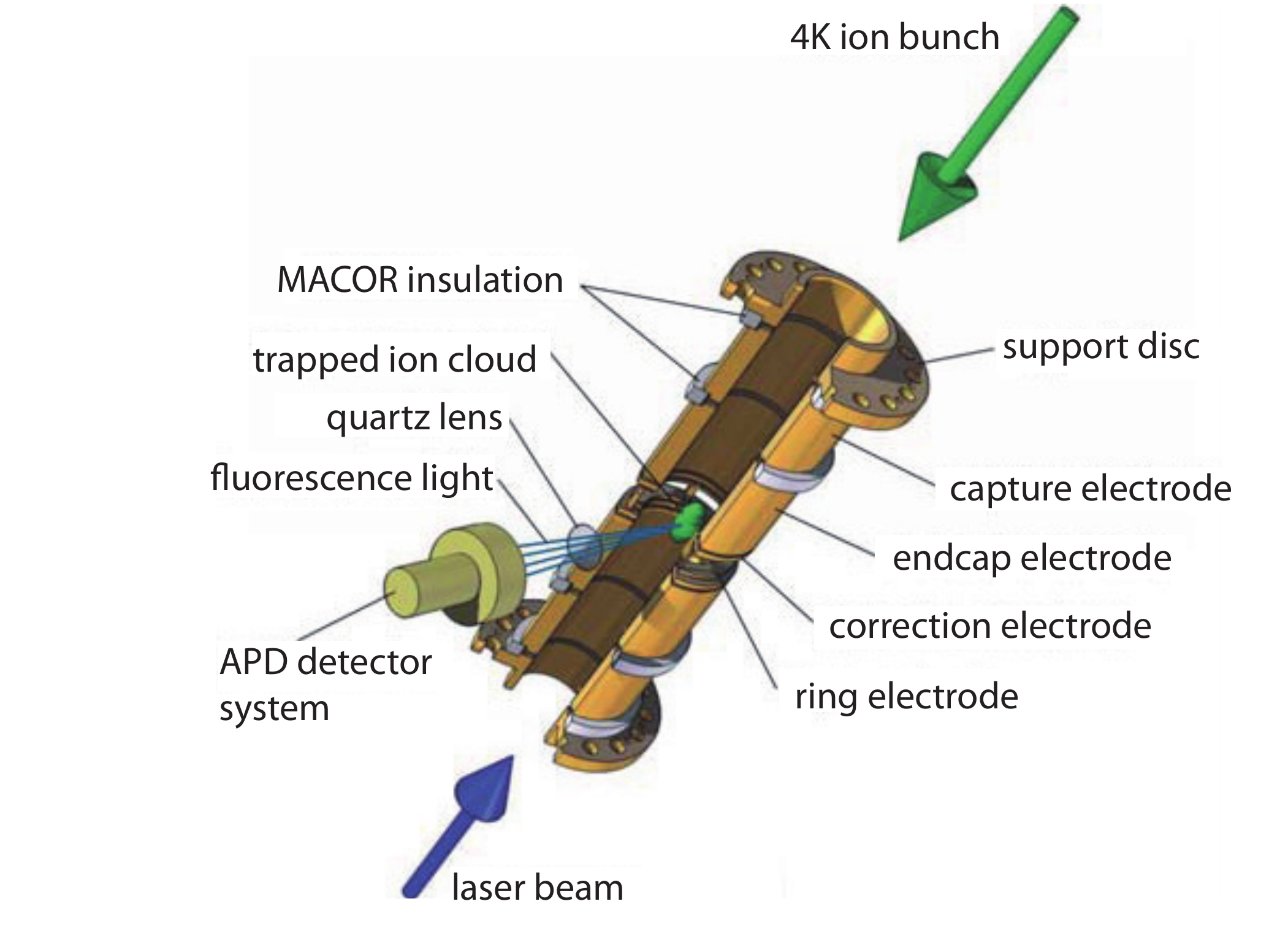}
\caption[SpecTrap setup scheme]{Illustration of the Penning trap used in the SPECTRAP experiment \cite{And10}.}
\label{fig::trap}
\end{figure}

The aim of the SPECTRAP experiment at GSI is to test QED in strong electric and magnetic fields~\cite{And10}. This is achieved by measuring the M1 hyperfine transition in the 2$s$-state in lithium-like and the 1$s$-state in hydrogen-like heavy ions of the same isotope by means of laser spectroscopy, and comparing the results with theoretical predictions~\cite{Sha01}. The highly charged ions (HCI) for these experiments will be delivered by the deceleration facility HITRAP~\cite{Qui01} and will be stored inside a penning trap where they are cooled using the resistive cooling technique. The trap is located inside the magnetic field of a pair of superconducting solenoids in Helmholtz configuration (see Fig.~\ref{fig::setup}) and cooled with liquid helium to reach a 
vacuum in the 10$^{-13}$~mbar range. The stored ions are excited by a tunable laser. The trap and the surrounding magnet setup contain four optical viewports which are oriented like a cross. Two of these viewports, perpendicular to each other, are equipped with spherical lenses next to the Penning trap electrode, to collimate the light originating from the trap center for transport through the beam tube towards the photodetector. To increase the photon flux at these two viewports, the respective trap openings on the opposite side are equipped with spherical mirrors that have their focal point in the trap center. Hence, light coming from the trap center is reflected back into the center. This should ideally double the light yield at the optical exits equipped with detectors (not taking into account imperfections of the mirrors). Figure~\ref{fig::trap} provides an overview of the Penning trap located inside the 
two solenoid magnets. Further details about the SPECTRAP experiment can be found in references \cite{And10} and \cite{Vog05}.\\
The Hyperfine transitions of the ions to be investigated span a wide wavelength range from the UV (240~nm) to the near infrared (up to 1600~nm) as listed in Table~\ref{tab::photonrates}. Since silicon Avalanche Photodiodes (APDs) are sensitive from about 300~nm up to 1100~nm with high quantum efficiencies (QE) of about 75\% in the peak region and still about 20\% at 1064~nm~\cite{RMD03} they are a promising detector candidate to cover a large part of this wavelength region and especially for the detection of the resonance in $^{207}\textrm{Pb}^{81+}$ at 1020~nm. Another advantage of APDs, compared to photomultipliers that are conventionally used for single-photon detection, is their insensitivity to magnetic fields.
At room temperature, APDs have been operated in fields up to 7.9~T without an effect on the observed amplification~\cite{Mar00}.
For cryogenically cooled APDs it has been found that at field strengths above 1~T there is some distortion of the APD signals~\cite{Gen11}. At the SPECTRAP experiment, the APDs will be mounted at a position where the calculated field strength due to the superconducting magnet of the trap is below 100~Gauss. We therefore do not expect any negative effects on APD performance.\\
Simulations~\cite{Ham08} have shown that the rates of fluorescence photons expected for transitions in the infrared region at the optical exits of the SPECTRAP setup
\begin{table}[h]
 \centering
 \caption[Count Rate Estimates]{Estimates of the expected photon rates at the optical exits of the SPECTRAP setup for some of the hyperfine transitions of interest~\cite{Ham08}.} 
 \begin{tabular}{|c|c|c|c|}
   \hline
   isotope & wavelength & lifetime & photon rate \\
   \hline
   $^{209}\textrm{Bi}^{82+}$ & 244 nm  & 0.4 ms & (625$\pm$225) kHz \\
   $^{207}\textrm{Pb}^{81+}$ & 1020 nm & 52 ms  & (6.5$\pm$2.1) kHz \\
   $^{209}\textrm{Bi}^{80+}$ & 1555 nm & 104 ms & (3.4$\pm$1.0) kHz \\
   \hline
  \end{tabular}
  \label{tab::photonrates}
\end{table}
are of the order of a few kHz only (see Tab.~\ref{tab::photonrates}). This results from the relatively long lifetimes of these hyperfine states in the order of a few milliseconds and from a very small solid angle accessible for fluorescence photon detection (see also Fig.~\ref{fig::trap}). Therefore detectors with single-photon counting capabilities and low dark count rates are required.\\
In this paper we report on our investigations about the applicability of type S0223 APDs manufactured by Radiation Monitoring Devices (RMD) for this purpose. These APDs have an active area of $2\times2$~mm$^2$, and are biased with up to 1.8~kV depending on their temperature. They feature a gain of up to 2000 at room temperature albeit with a high dark current. In order to reach higher gains and decrease thermal noise, we operate the diodes at cryogenic (liquid nitrogen) temperatures.
\section{Setup for APD characterization}\label{sec::setup}
\begin{figure}[htbp!]
\centering
\includegraphics[width=\textwidth]{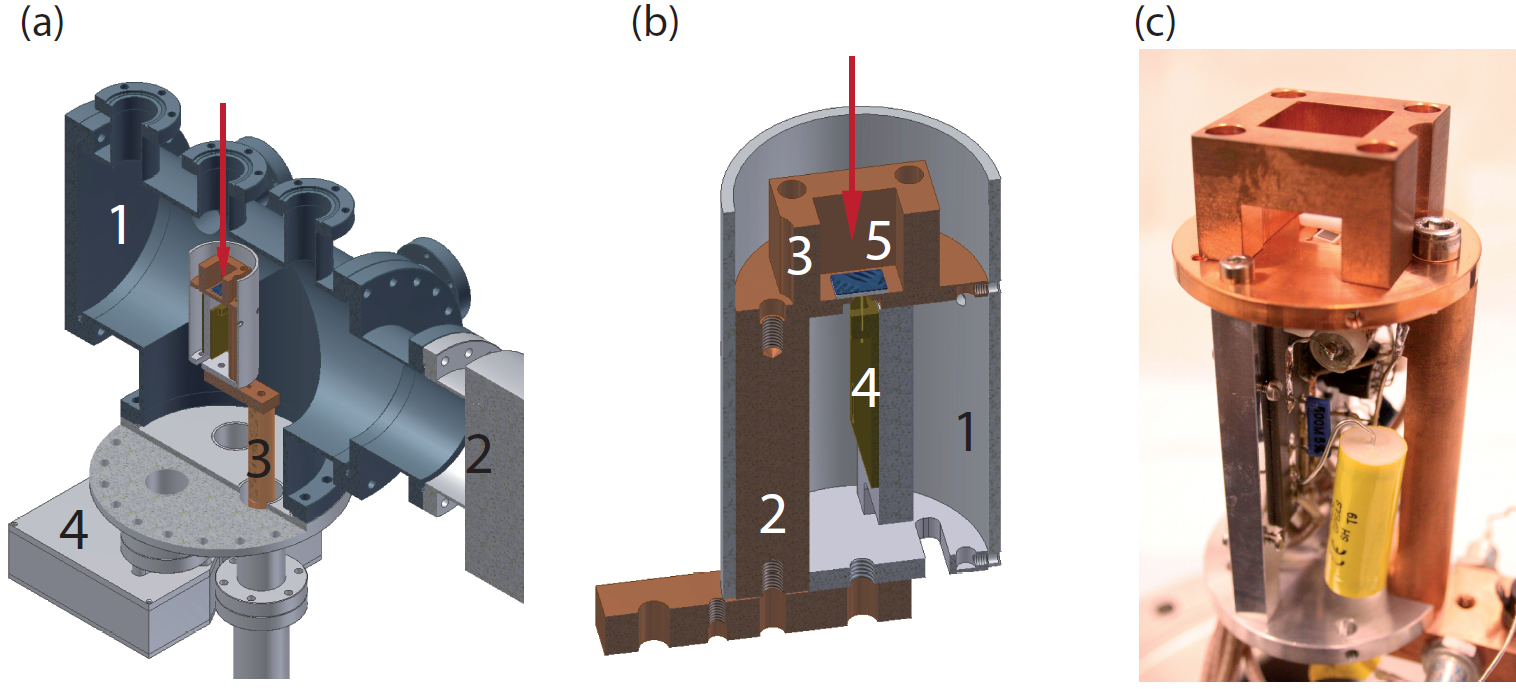}
\caption[Setup scheme]{(a) Vacuum setup: 1 - vacuum chamber, 2 - turbomolecular pump, 3 - cold finger, 4 - electronics housing, e.g. for preamp main stage. 
(b) Detector mount: 1 - electric shielding, 2 - intersection and heat conductor, 3 - cold shield, 4 - first preamplifier stage, 5 - APD. The red arrows indicate the direction of incident light.
(c) Photograph of the detector mount without electric shielding.}
\label{fig::vacuum_chamber}
\end{figure}

\begin{figure}[h]
\centering
\includegraphics[width=0.4\textwidth]{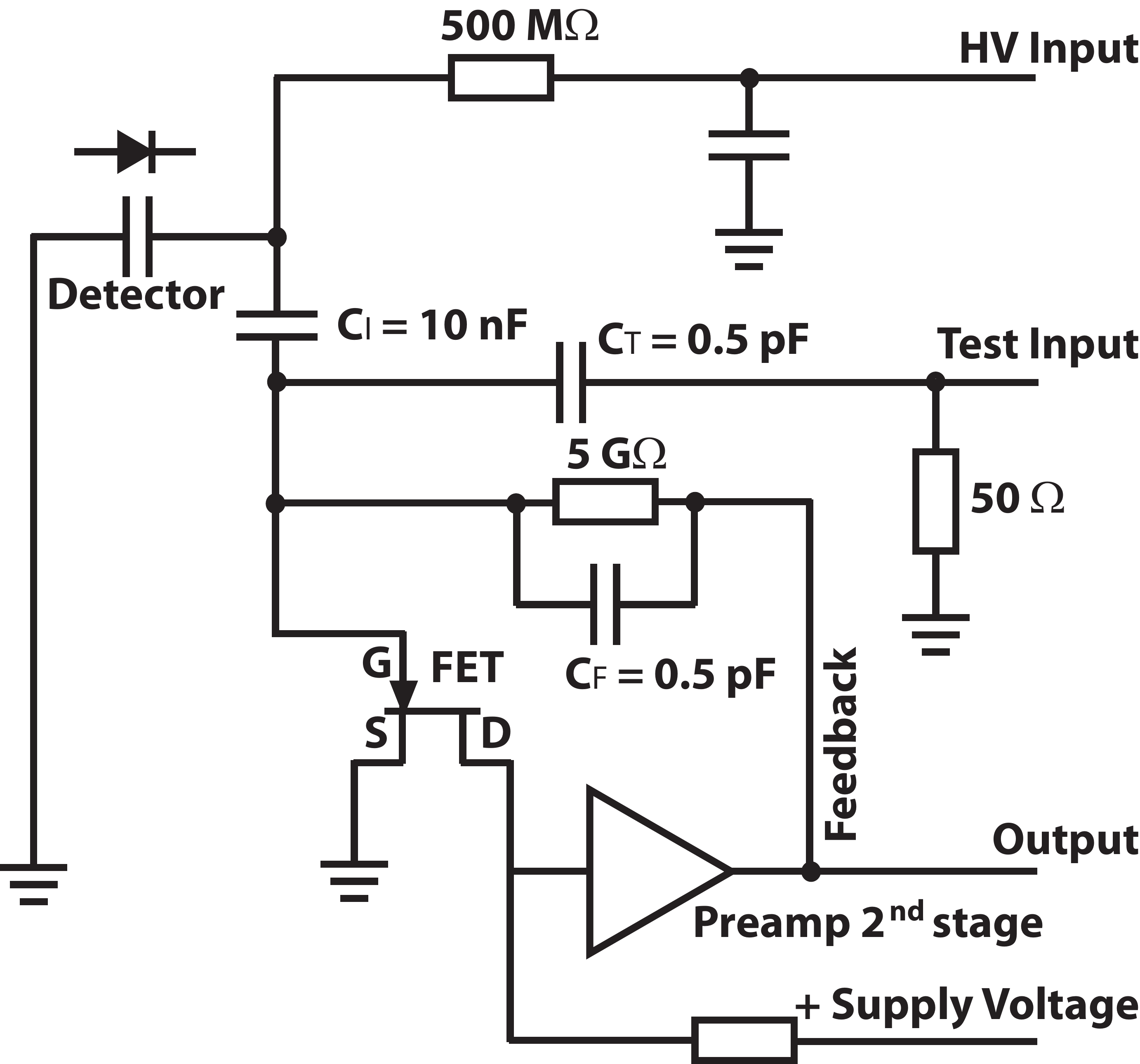}
\caption[Preamp scheme]{Scheme of the preamplifier. The single FET represents the parallel U430 JFET. The triangle is a symbol for the second stage of the preamplifier.}
\label{fig::preampscheme}
\end{figure}

In order to minimize thermal noise of the APDs the devices are cooled down to near liquid-nitrogen temperatures. To avoid condensation of gas molecules on the detector, the APDs are mounted and operated inside a small vacuum chamber 
(see Fig.~\ref{fig::vacuum_chamber}) at a pressure of 3$\cdot$10$^{-7}$~mbar. The vacuum system consists of a Pfeiffer MVP~040-2 
diaphragm pump and a Leybold TW~300 turbomolecular pump. A special mounting structure as shown in Fig.~\ref{fig::vacuum_chamber}b inside the vacuum chamber is used to hold the APD and the first stage of the preamplifier board and to provide electric shielding. This structure is mounted on a copper cold finger that is inserted into a liquid nitrogen dewar to cool the components. 
With this setup a minimum temperature of -178$^{\circ}$C was reached at the copper detector mount. The intersection between the cold finger and the detector mount is equipped with two Zener-diodes (not shown) which serve as heaters and can be used for temperature control.
For temperature readout PT~1000 sensors are located near the heaters and on the upper copper plate of the detector mount next to the APD. The heating can also be used to moderately bake out the system. It is possible to operate the setup at any temperature between -178$^{\circ}$C and about +100$^{\circ}$C with a precision of $\pm$0.5$^{\circ}$C using a LabVIEW based PID temperature control program. To avoid damage to the APD, the temperature control ensures a maximum temperature gradient of 2$^{\circ}$C/min at the detector while cooling down. The LabVIEW program is also used to set and monitor the bias voltage and to monitor the current drawn by the APD by remotely controlling an ISEG NHQ~224M high voltage power supply. That way complete measurements can be done automatically changing the bias voltage and taking a spectrum at each voltage with an Ortec 926-M32-USB multi channel analyzer.\\
The APD is connected directly to the first stage of a low noise preamplifier. A small teflon piece is used to press the APD onto the copper detector mount in order to improve the thermal connection and to tighten the APD to suppress microphonic effects. The circuitry of the low noise preamplifier board, shown in Fig.~\ref{fig::preampscheme}, is based on a layout used in the Mainz Neutrino Mass 
Experiment~\cite{Wei92}. This charge sensitive device is divided into two stages. The first amplification stage consists of two matched Vichay U430 n-channel JFET transistors assembled in a TO-48 package~\cite{Vis01} which are connected in parallel to reduce the noise. The second stage is mounted outside the vacuum in a special aluminum case directly connected to a CF~35 vacuum feedthrough. Another CF~35 feedthrough with aluminum case is used for temperature readout and heating connections. The main amplifier used in the setup is a CAEN N968 Spectroscopy Amplifier. Further electronics settings and signal processing is described in the next sections as needed.\\
A Tektronix AFG~3102 function generator is used to generate an electronic test pulse which is connected to the test input of the preamplifier. This pulse simulates a $E=100$~keV signal for calibration purposes. Since the mean energy for creating an electron-hole pair in silicon is $W=3.76$~eV at 77~K~\cite{Ber68}, the charge that has to be deposited on the capacitance C$_T$ is given by
\begin{equation}
 Q=\frac{E}{W}=\frac{1\cdot10^{5}\textrm{ eV}}{3.76\textrm{ eV}}\cdot e=4.3\cdot10^{-15}\textrm{ C},
\end{equation}
where $e$ is the elementary charge.
Thus the amplitude of the pulse loading the 0.5~pF capacitance at the test input has to be
\begin{equation}
 U=\frac{Q}{C_T}=\frac{4.3\cdot10^{-15}\textrm{ C}}{0.5\cdot10^{-12}\textrm{ F}}=8.6\textrm{ mV}.
\end{equation}
To generate this voltage the function generator is set to a pulse amplitude of 1~V and the signal is reduced to 8.6~mV by two ($20\pm1$)~dB attenuators to allow a stable operation. With 
\begin{table}[htbp]
 \centering
  \caption[Pulse settings]{Settings of the function generator for LED and Test pulses to determine gain and noise of the APD.}
  \begin{tabular}{|c|c|c|}
   \hline
   & Test pulse & LED pulse \\
   \hline
   rise time (ns) & 5 & 5 \\
   decay const. ($\mu$s) & 40 & 40 \\
   frequency (kHz) & 0.8 & 1.6 \\
   amplitude (V) & 1 & 1.1 - 0.8 \\
   \hline
  \end{tabular}
  \label{tab::pulsersettings}
\end{table}
this reference pulse the noise contribution of the preamplifier electronics without an attached detector has been determined to be $\sigma = 281$~eV (corresponding to 75 electrons).\\
To induce detector signals the light of a pulsed red LED (628~nm) is coupled onto the detector via an optical fibre. The LED is also driven by the AFG 3102 function generator. The pulse settings used in the measurements to determine gain and noise of the APDs are listed in table \ref{tab::pulsersettings} in this section.
\section{APD characterization}
\label{sec::APDs}
\begin{figure}[htbp!]
\centering
\includegraphics[height=0.6\textwidth, angle=-90]{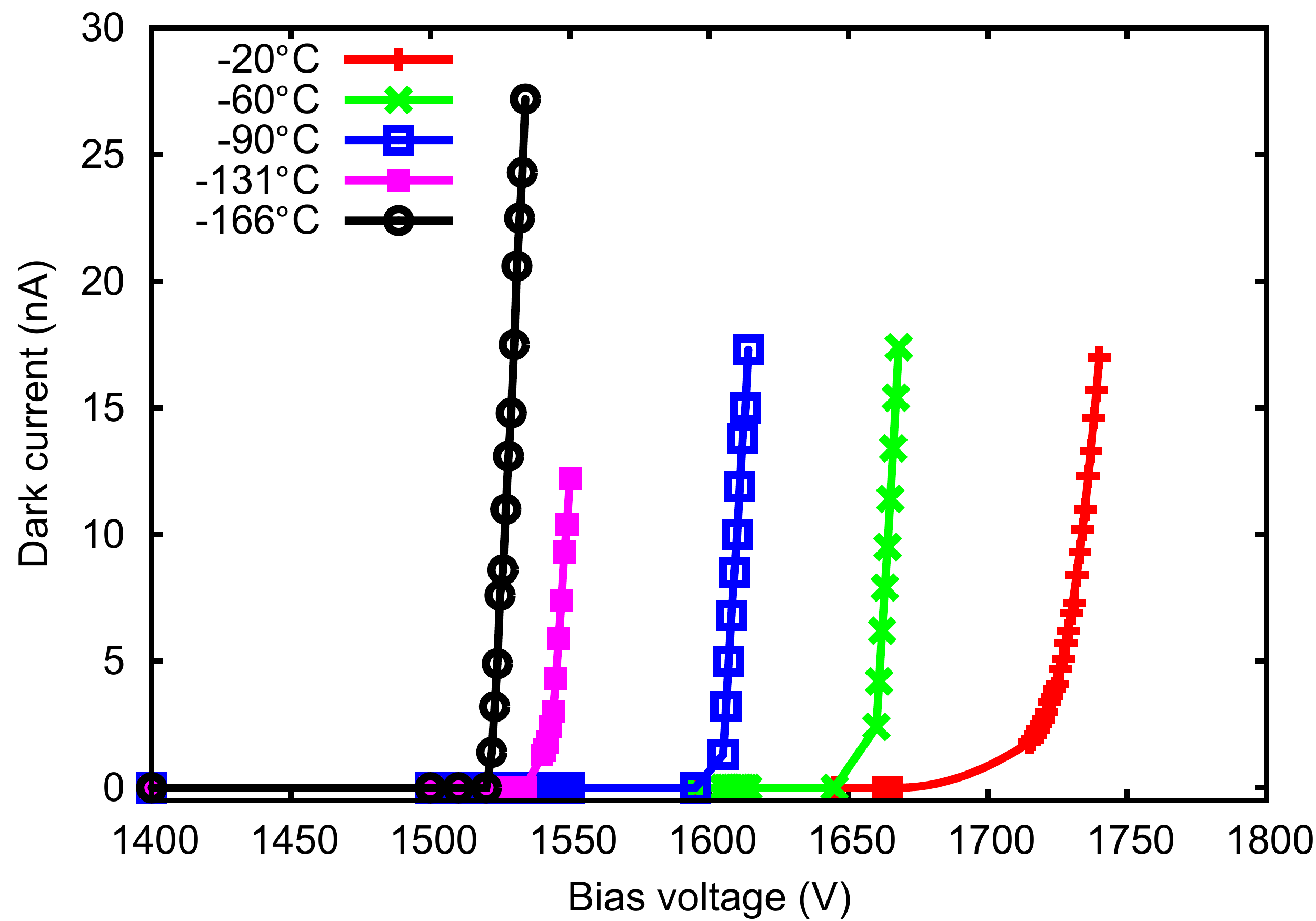}
\caption[S0223 dark current]{Typical dark current behavior of a S0223 APD at different temperatures}
\label{fig::darkcurrent}
\end{figure}

The results of measurements performed to characterize the general behavior of the APDs (dark current, relative gain and noise) at different ambient temperatures are presented in the following section.
\paragraph{Dark current}
\label{sec::darkcurrent}
\nl
The APD dark current is measured using an amperemeter implemented in the ISEG NHQ~224M high voltage supply that provides a resolution of 100~pA. Fig.~\ref{fig::darkcurrent} shows the dark current as a function of bias voltage 
at different APD temperatures. It is obvious that with decreasing temperature the breakdown voltage of the APD also decreases. At the same time, the slope of the dark current becomes much steeper when approaching the breakdown voltage. When cooling the APD to low temperatures, this behaviour allows to operate the diode very close to the breakdown voltage and thus achieve high gain factors.
\paragraph{Gain}
\label{sec::gain}
\nl
For the APD gain measurements, we induce signals by shining light from a pulsed LED onto the diode. In the electronics setup used for this measurement the preamplifier signals are then further amplified and shaped by a CAEN Mod. N968 spectroscopy amplifier. The optimum shaping time was determined to be 2~$\mu$s. The amplified signals are finally recorded using an ORTEC 926-M32-USB multichannel analyzer.\\
To determine the relative APD gain at a certain bias voltage, we take the ratio of the signal amplitude at this voltage to the signal amplitude at 10~V bias, where it can safely be assumed that no avalanche multiplication takes place. As, at low voltages, there will also be recombination effects, we can not determine absolute gains with this method. For the latter, one needs to know at precisely which voltage the effects of recombination and the onset of amplification of the APD cancel each 
\begin{figure}[h]
\centering
\includegraphics[height=0.6\textwidth, angle=-90]{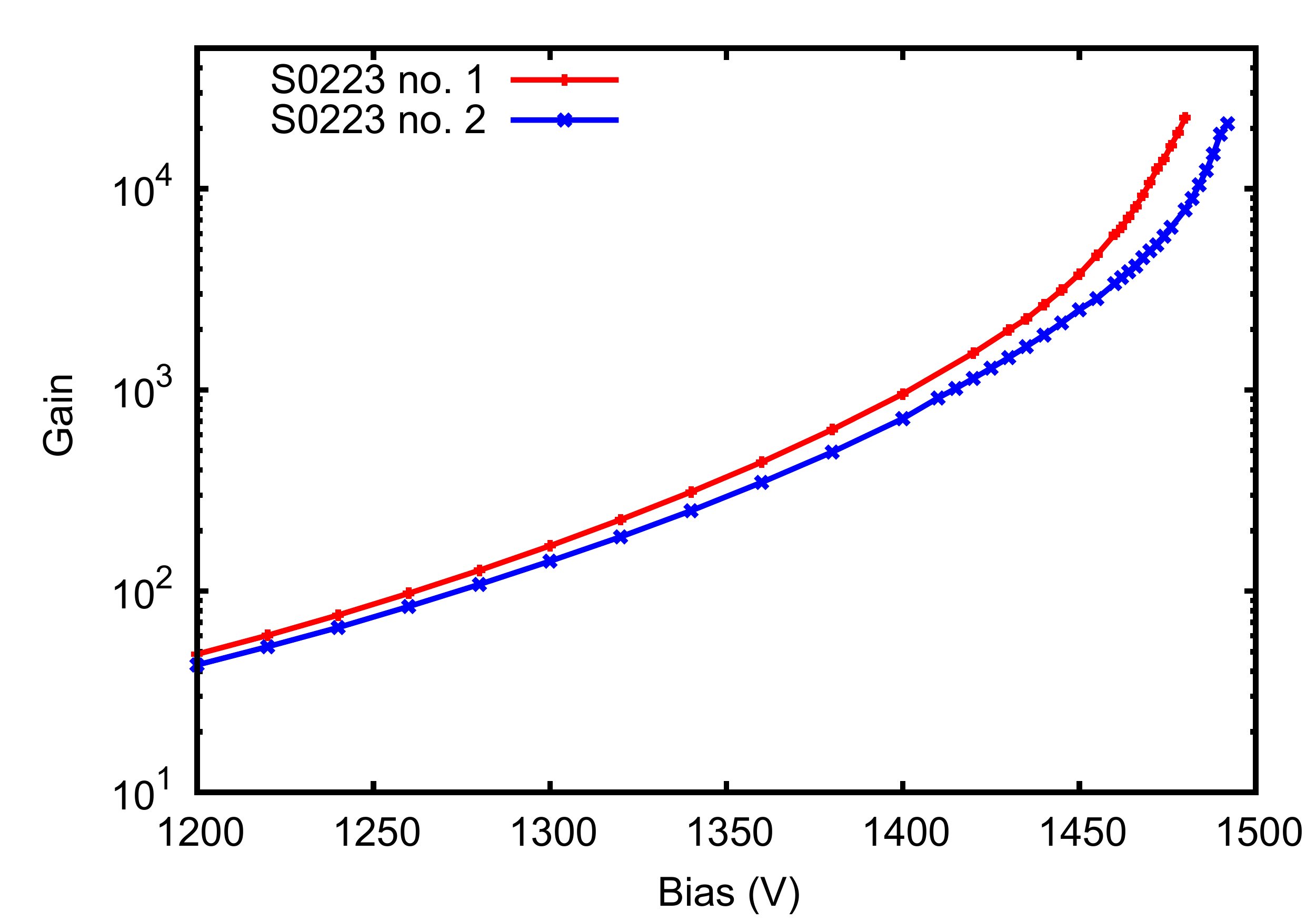}
\caption[S0223 Gain]{Relative gain of two different S0223 type APDs}
\label{fig::gain}
\end{figure}
other. This point will be subject of further investigations.\\
If, at higher gains, the signal amplitude starts to exceed the dynamic range of the ADC, the amplitude of the pulse driving the LED is lowered. The ratio between LED pulse positions in the spectra at two different amplitudes but equal bias voltage is used as correction factor in the calculation of the gain.
Fig.~\ref{fig::gain} shows the results of the gain measurement for two different S0223 type APDs, with the low noise setup and cryogenic cooling to about -176$^{\circ}$C. Relative gains of $M \geq 2\cdot10^4$ were observed.
%
\paragraph{APD operation bias}
\nl 
The bias voltage used to operate the APDs in the following measurements was determined by the following considerations:
\begin{itemize}
 \item The operation bias should not be chosen too close to breakdown since the breakthrough voltage depends on the APD temperature and small temperature drifts should not be able to shift the operation bias beyond this voltage.
 \item On the other hand the operation bias should be close to the breakthrough voltage in order to be able to use high gain values for the single-photon measurements.
\end{itemize}
We found that a bias voltage of 1480~V is a reasonable compromise between these two conflicting requirements, with a distance of about 10~V to the breakthrough voltage at -176$^{\circ}$C but still a very high relative gain of $M\approx2.2\cdot10^4$. During the measurements presented in the following this choice has allowed for stable operation of the setup for several weeks.\\
The value of the breakdown voltage varies between different APDs of the same type by some 10~V. At room temperature values between 1840~V and 1870~V are stated in the datasheet of the procured APDs. The breakdown voltages decreases with temperature by about 1.4~V/$^{\circ}$C.
\section{Single-photon calibration and counting}\label{sec::singlephotons}
\begin{figure}[h]
\centering
\includegraphics[height=7.5cm,angle=-90]{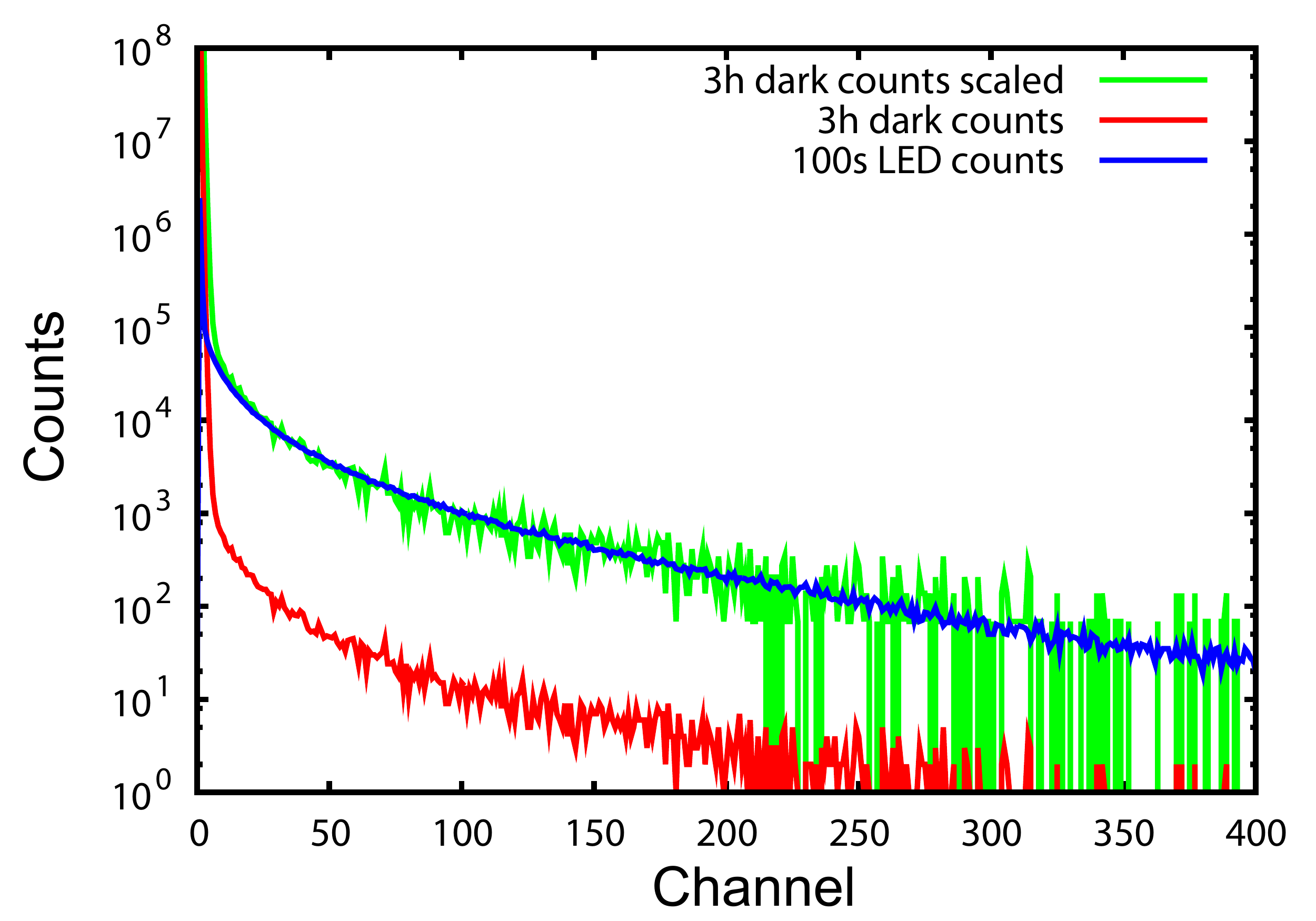}
\includegraphics[height=7.5cm,angle=-90]{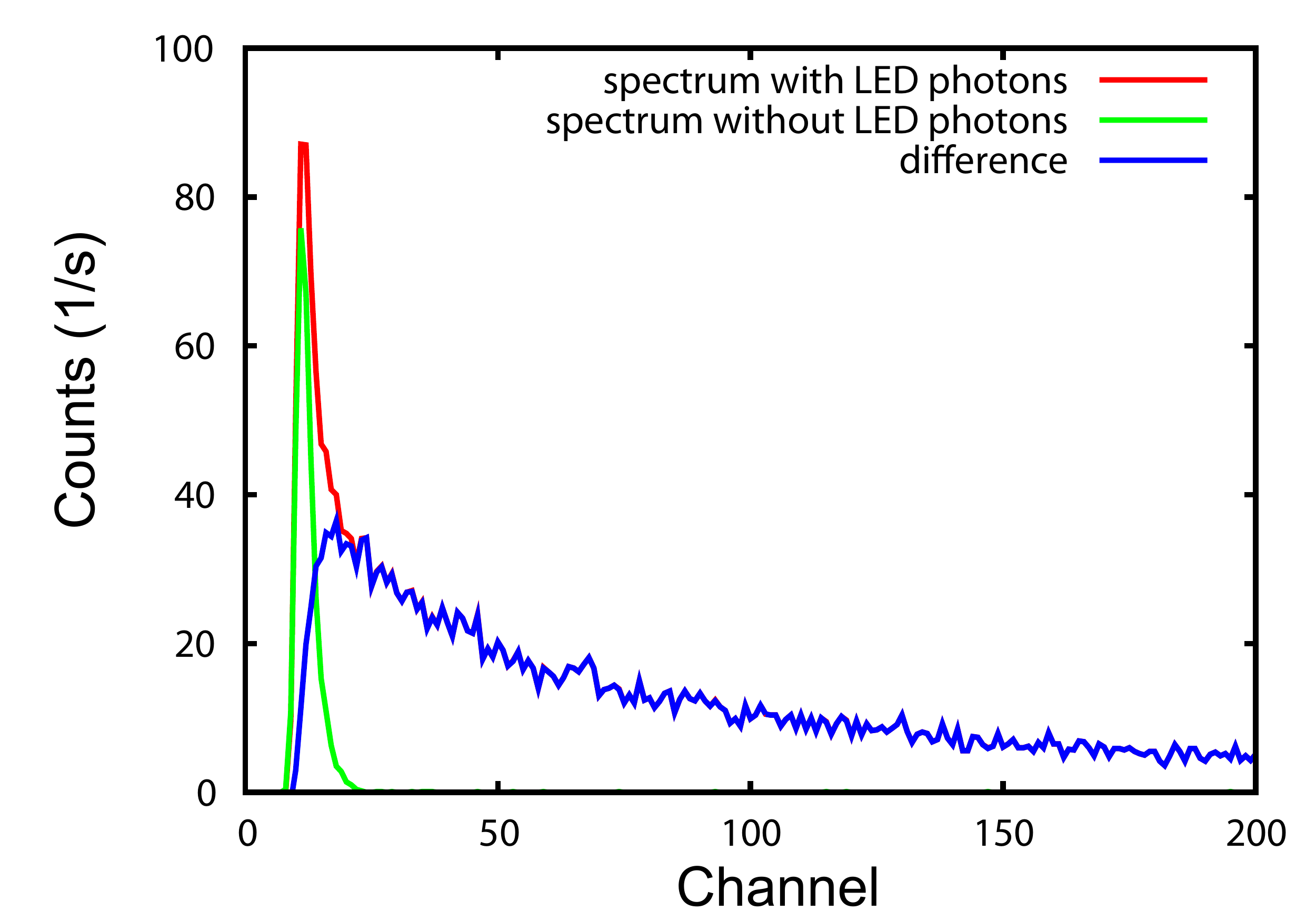}
\caption[Single-Photon spectrum]{Left: APD spectra taken with and without illumination by single-photons generated with the pulsed LED. The amplitude distribution of the scaled dark count spectrum matches the shape of the single-photon events.
Right: amplitude spectrum at 1480~V bias with and without single-photon source together with the difference spectrum showing only photon events.}
\label{fig::spectrumcomparison}
\end{figure}

\begin{figure}[htbp!]
\centering
\includegraphics[width=\textwidth]{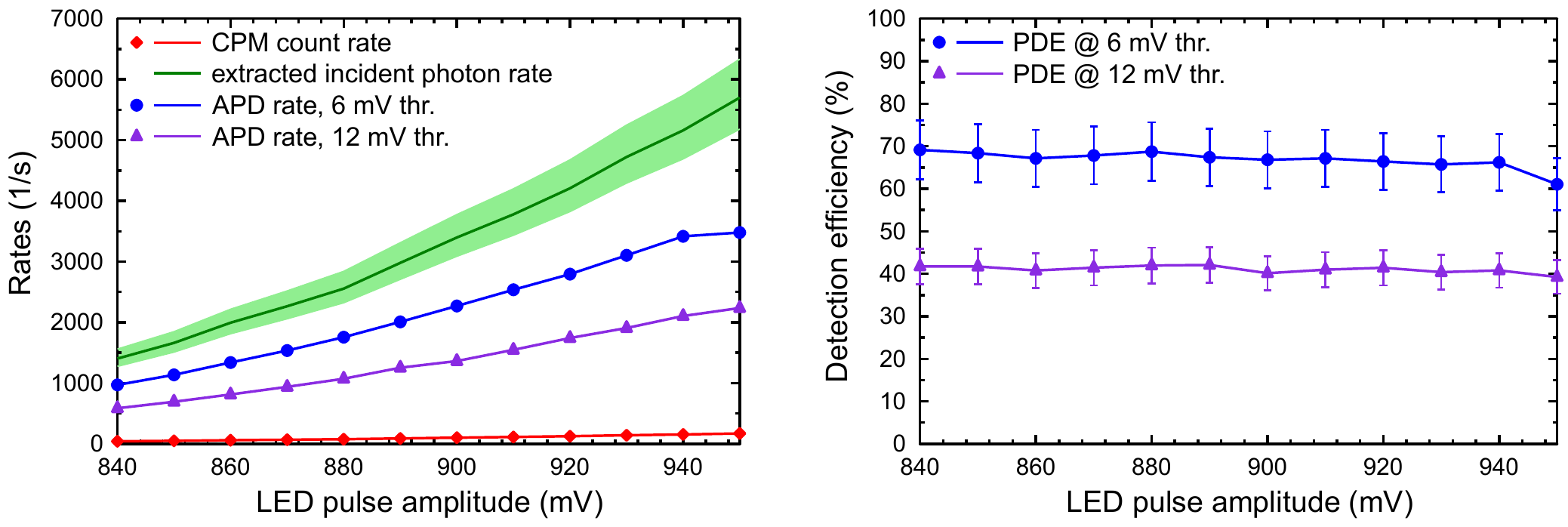}
\caption[Single-Photon measurement]{The CPM count rates for different LED pulse amplitudes are shown in red. With a CPM QE of $(3\pm0.3)$\% the number of incident photons is represented by the green band. Comparing the blue and purple datapoints representing the APD count rates at 6~mV and 12~mV discriminator threshold, respectively, with the incident band, the APD has an average photo detection efficiency of $(67\pm7)$\% in the case of 6~mV threshold and $(41\pm4)$\% in the case of 12~mV threshold.}
\label{fig::singlephotons}
\end{figure}

\paragraph{Single-photon light source}
\label{sec::lightsource}
\nl
To determine the photo detection efficiency of the APDs, a single-photon light source has been realized using a red LED (628~nm) that is operated by applying short (40~ns) pulses with amplitudes between 840~mV and 950~mV 
\begin{table}[b]
 \centering
   \caption[Single-Photon pulse settings]{Settings of the function generator to generate LED pulses that result in mainly single-photon events incident on the APD.}
  \begin{tabular}{|c|c|}
   \hline
   Single-Photon pulse settings & \\
   \hline
   rise time (ns) & 5 \\
   fall time (ns) & 5 \\
   width (ns) & 40 \\
   frequency (kHz) & 100 \\
   amplitude (mV) & 840 - 950 \\
   \hline
  \end{tabular}
  \label{tab::singlephotonsettings}
\end{table}
(see table~\ref{tab::singlephotonsettings}). 
The number of photons emitted by the LED into the solid angle covered by the APD is Poissonian distributed. The possibility of two photons emitted simultaneously into this solid angle becomes negligible for a suitable choice of the expectation value $\mu$ of the Poissonian distribution:
\begin{equation}
 P_{\mu}(n)=\frac{\mu^n}{n!}e^{-\mu}.
\end{equation}
E.g. for $\mu=0.1$ the possibility that no photon is emitted into the detector's solid angle is $P(0)=0.905$. The possibility for one photon hitting the detector is $P(1)=0.090$ and for two photons $P(2)=0.005$. The count rate of the detector can be described by
\begin{equation}
  N = N_P + D,
\end{equation}
with
\begin{equation}
  N_P = r \cdot \sum_{n=1}^{\infty} \left[ \varepsilon_n \cdot P_\mu(n) \right],
\end{equation}
where $N$ is the total count rate, $N_P$ the count rate due to the light source, $D$ the dark count rate and $r$ the rate of LED pulses. $\varepsilon_n$ is the probability for $n$  photons, simultaneously hitting the device, to be detected. To calculate $\varepsilon_n$ we have to consider how many photoelectrons are produced by the incident photons, and what the chances are for these photoelectrons to contribute to a measurable signal. The number of photoelectrons $k$ produced by $n$ incident photons follows a binomial distribution $f(k;n,QE)$ that depends on the quantum efficiency (QE) of the detector. If we label the efficiency for detection of a signal produced by $k$ photoelectrons by $\varepsilon_{e,k}$ we obtain for $\varepsilon_n$
\begin{equation}
  \varepsilon_n = \sum_{k=1}^{n} \varepsilon_{e,k} \cdot f(k;n,QE)
\end{equation}
The probability $\varepsilon_{e,k}$ to actually detect a certain number of photoelectrons produced in the detector depends on the collection efficiency and the amplification of the device. To calibrate our single-photon source we use a channel photomultiplier (CPM) manufactured by PerkinElmer \cite{Per1}. These devices have a collection efficiency larger than $95\%$ and a high amplification of the order $1\cdot 10^7$ that makes them ideal for photon counting. In this case, we can take a shortcut in calculating $\varepsilon_n$ by assuming that each event, where at least one photoelectron is produced, is detected, i.e. $\varepsilon_{e,k}({\rm CPM}) \approx 1$. The probability that no photoelectron is produced from $n$ photons is given by
\begin{equation}
  f(0;n,QE) = (1-QE)^n
\end{equation}
Therefore, the probability that at least one photoelectron is produced and subsequently detected is in our case given by
\begin{equation}
  \varepsilon_n({\rm CPM}) \approx 1 - (1-QE)^n
\end{equation}
and thus
\begin{equation}
  N_P({\rm CPM}) \approx  r \cdot \sum_{n=1}^{\infty} \left[ (1 - (1-QE)^n) \cdot P_\mu(n) \;\right] .
\end{equation}
$N_P({\rm CPM})$ is therefore determined by the expectation value $\mu$, the QE of the detector and the rate $r$ of the pulse driving the light source.\\
Using short (40~ns) pulses with 850~mV amplitude and a pulse rate of $r=100$~kHz, we measure a signal rate of $N_P=50$. 
The active area (\o~=~15~mm) of the CPM has, for this measurement, been limited using an aperture of 2.2~mm, drilled into a black plastic cap which is plugged onto the CPM. The area of the opening is 5\% smaller than the active area of 4~mm$^2$ stated on the datasheet of the APD. Measurements of the sensitivity distribution of a larger APD manufactured by RMD showed, however, that the detection efficiency decreases close to the edges of the specified active area~\cite{Sha09}, such that we do not expect a significant systematic error due to the small difference in the opening area.
With a quantum efficiency of 3\% of the CPM at 628~nm, this result corresponds to $\mu = 0.017$, with $P(0)=0.983$ and $P(1)=0.016$. This means that every 63'rd pulse of the LED emitts a single-photon into the solid angle of the detector, while the probability of two or more photons to hit the detector is with $P(2)=1.37\cdot10^{-4}$ negligible.\\
An additional method to prove that our light source emits single-photons into the solid angle of our detector is to compare a long term amplitude spectrum taken without LED light to a spectrum with illumination. Without illumination, electron hole pairs are generated in the diode by thermal excitation of electrons into the conduction band. These thermally generated electrons cannot be distinguished from electron hole pairs generated via the photo-electric effect. A dark count spectrum should, therefore, after a certain time have the same amplitude distribution as a spectrum of the single-photon source. This is demonstrated in Fig.~\ref{fig::spectrumcomparison}, left, that shows a comparision between a single-photon 
spectrum and a scaled version of the dark count distribution obtained in a 3~h measurement. In Figure~\ref{fig::spectrumcomparison}, right, the amplitude distribution of single-photon signals at 1480~V bias is extracted by subtracting a dark count spectrum taken with the same measurement lifetime.
\paragraph{Single-photon detection efficiency at 628~nm}\label{sec::628nm}
\nl
Due to recombination not every primary electron hole pair will trigger a detector signal. The photo detection efficiency is therefore given by the ratio of the background corrected detector signal rate to the number of incident photons and is less than or at best equal to the QE (which is the ratio of generated electron hole pairs to the number of incident photons). In the following the photo detection efficiency of the APDs will be discussed.\\
From the measurements with the channel photomultiplier, described above, we can extract the rate of photons incident on our APD at different LED pulse amplitudes.
\begin{figure}[htbp]
    \subfloat[Dark count rate $D$ vs. threshold]{\includegraphics[width=7cm]{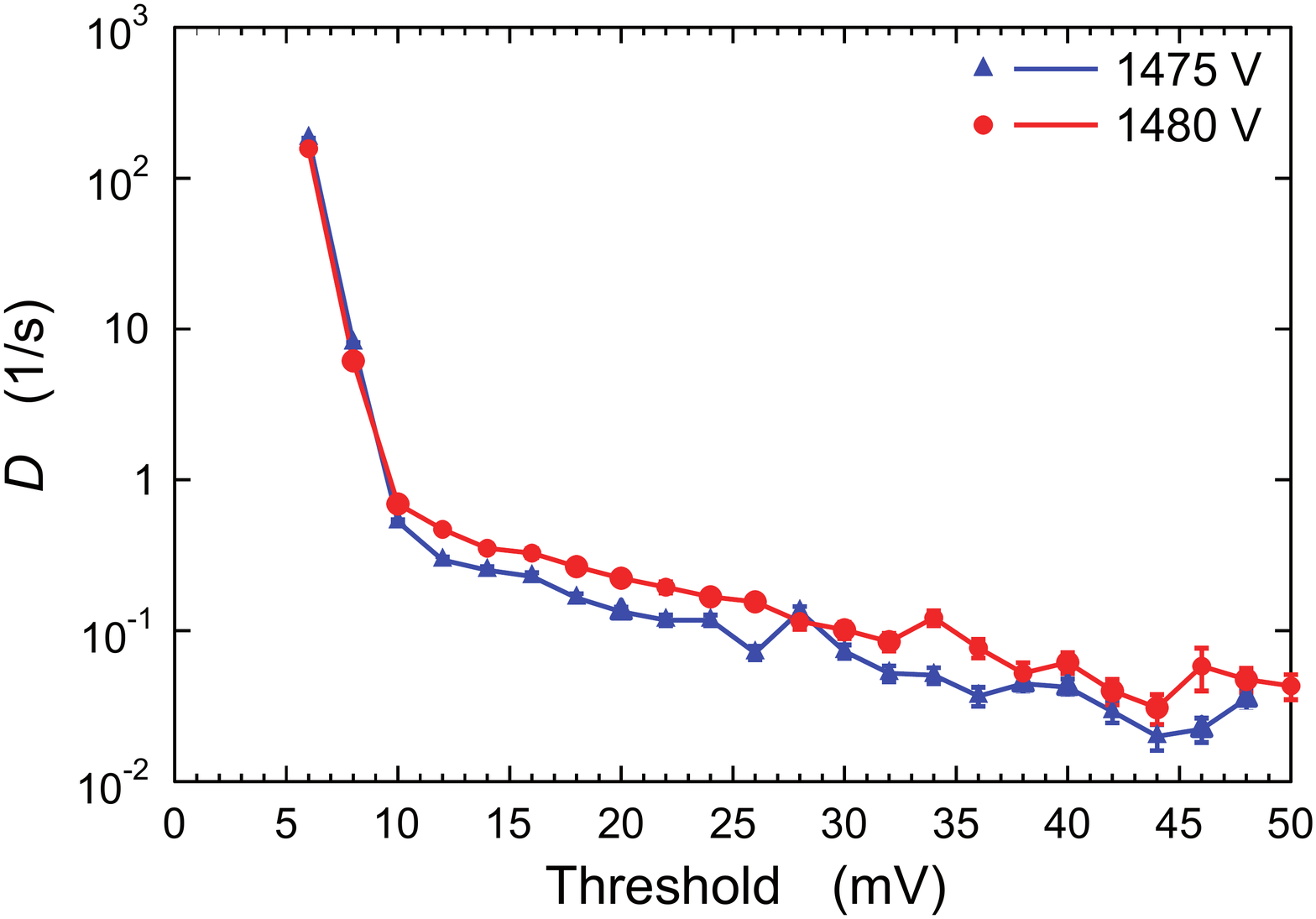}}
    \subfloat[Signal rate $S_{net}$ (background corrected) vs. threshold]{\includegraphics[width=7cm]{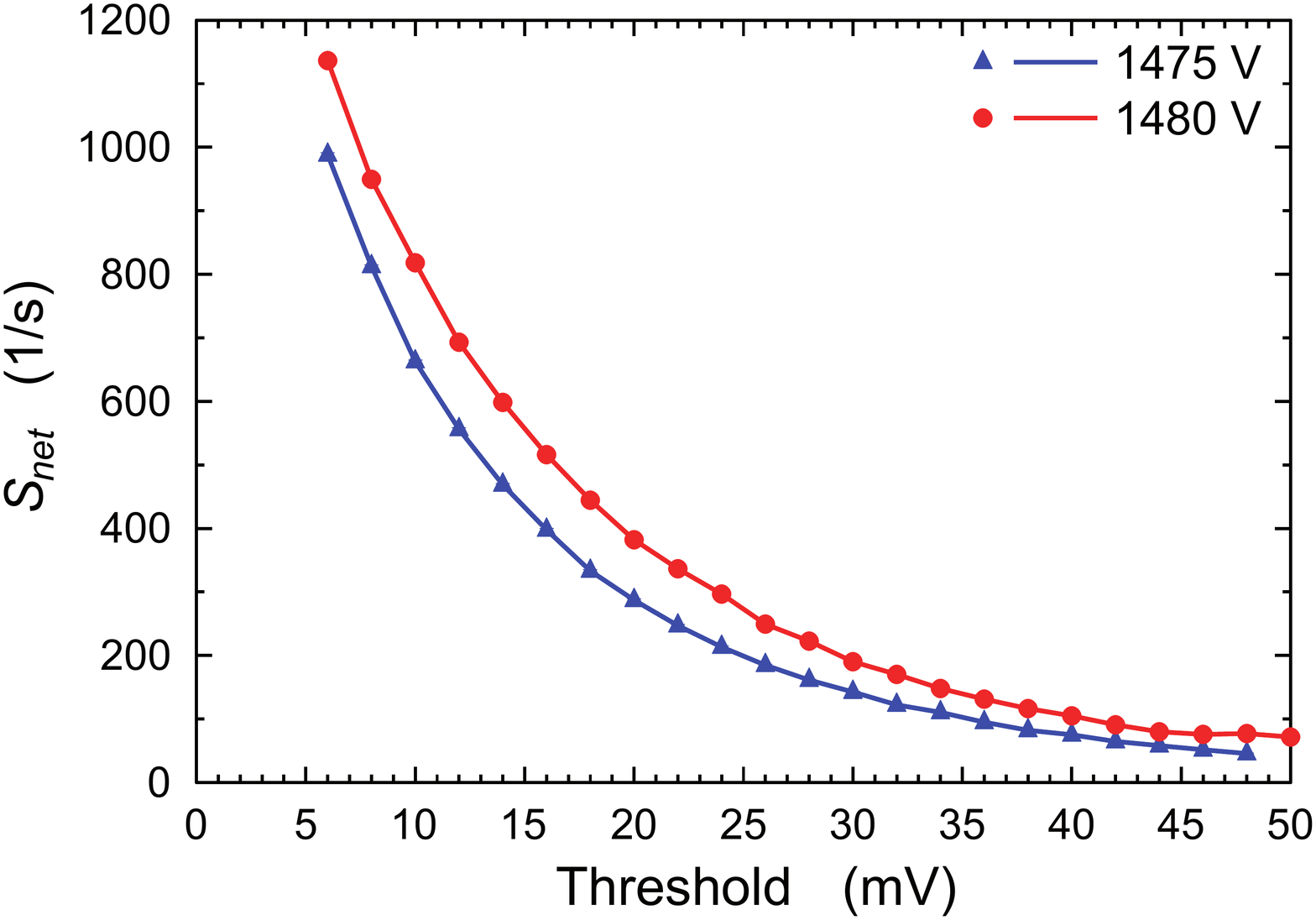}}\\
      \centering
    \subfloat[$S_{net}^2 / D$ vs. threshold]{\includegraphics[width=7cm]{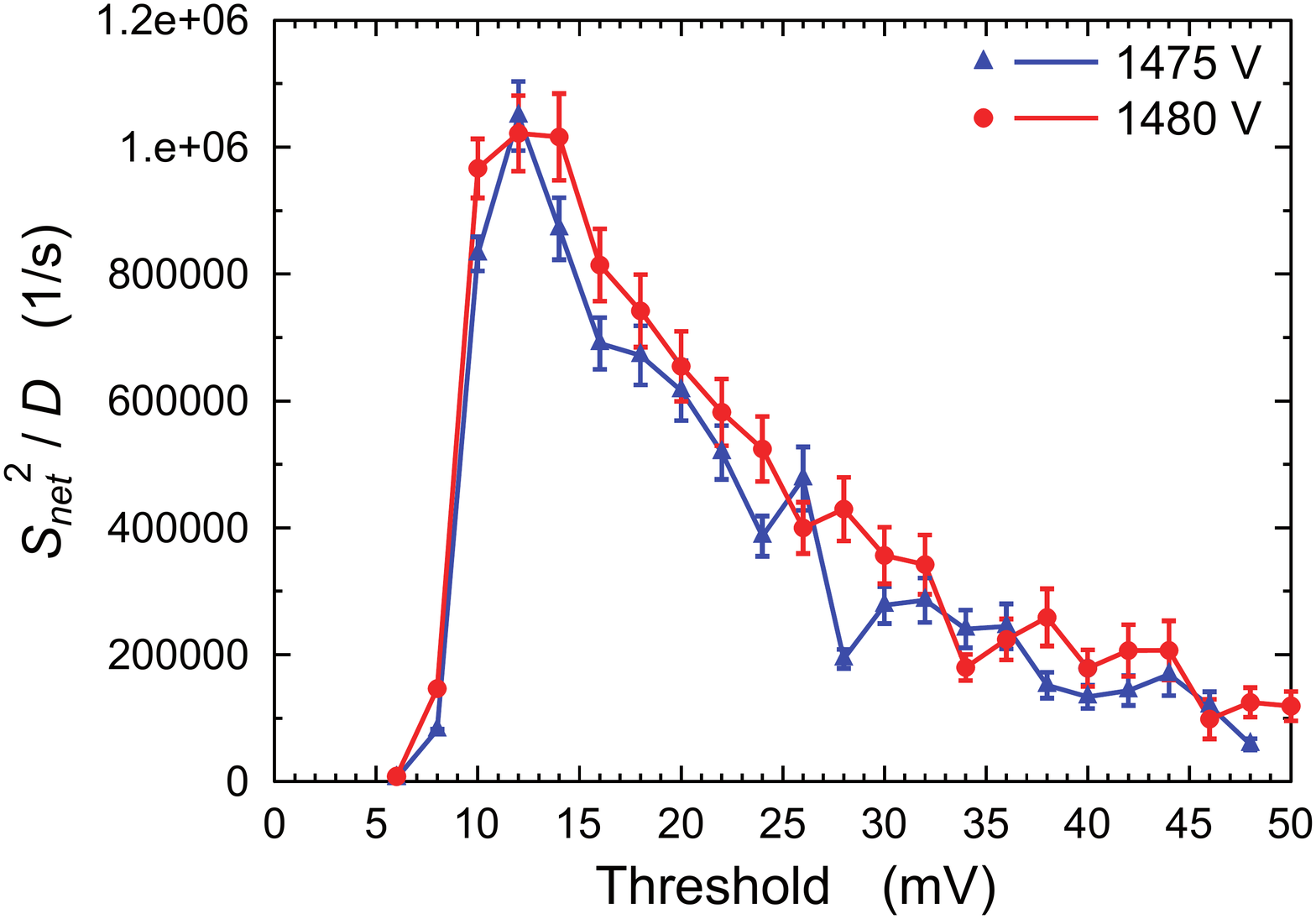}}
\caption{Comparison of the obtained values vs. threshold for two different bias voltages: a) Dark count rate $D$ , b) Signal rate $S_{net}$ (background corrected)  , c) $S_{net}^2 / D$}
\label{fig::oppoint}
\end{figure} 
To determine the APD count rates at a given signal amplitude, we discriminated the main amplifier signals using an ORTEC Mod. 550 single-channel analyzer. The output of the discriminator was connected to an ORTEC Mod. 416A gate generator, that was used to elongate the logic signals to avoid double counts from over/undershoots of the main amplifier signal. These signals were then counted by a SIS 3803 scaler module.\\
The count rates with and without LED light were taken for several discriminator threshold settings from 6~mV to 50~mV. When lowering the threshold, the background rate as well as the signal rate increases. In case the photon signal can be cleanly seperated from the background, the LED signal rate should remain constant when further decreasing the discriminator threshold and only the background rate should rise. In our case, where the signal peak is not clearly seperated from the background, the preferred discriminator threshold setting
cannot be determined easily. Since we do not want to loose too many single-photon signals the threshold has to be set to a value where the LED signal rate is highest for a reasonable dark count rate.
To choose the threshold settings one has to consider that in the end the aim is to optimize the measuring time per laser wavelength at the SPECTRAP experiment. In order to distinguish a signal from the background with a certain significance $n\sigma$ the number of detected photons $N_P\cdot t$ must be at least
\begin{equation}
  N_P \cdot t = S \cdot t \cdot \varepsilon = n \cdot \sqrt{D\cdot t},
\end{equation}
where $S$ is the rate of photons incident on the detector, $t$ the measuring time, $\varepsilon$ the detection efficiency of the detector and $D$ the dark count rate. This leads to a measuring time $t$ of at least
\begin{equation}\label{eqn::mintime}
  t = \frac{n^2}{\varepsilon^2} \cdot \frac{D}{S^2} = \frac{n^2 \cdot D}{S_{net}^2}
\end{equation}
with $S_{net}=S \cdot \varepsilon$. The measuring time is thus proportional to the ratio of the dark count rate and the background corrected signal rate $S_{net}$ squared. In other words, to minimize the measuring time we have to maximize:
\begin{equation}
  \frac{1}{t} \propto \frac{S_{net}^2}{D}.
\end{equation}
Figure~\ref{fig::oppoint} shows the dark count rate (upper left plot), the background corrected single-photon rate (upper right) and $\frac{S_{net}^2}{D}$ for the investigated RMD S0223 APD operated at 1475~V and 1480~V, respectively. To achieve the maximum photo detection efficiency the detector should therefore be operated at the minimum threshold of 6~mV, where the dark count rate is still at an acceptable value of $\approx230$~s$^{-1}$. To minimize the measurement time required for signal detection at the SPECTRAP setup we should on the other hand use a threshold of 12~mV with these APDs and under the described operating conditions.\\
For each threshold value the photo detection efficiency of the APD can be calculated by dividing the incident photon rate measured with the reference CPM by the background and deadtime corrected signal rate of the APD. The results obtained from these measurements are shown in Fig.~\ref{fig::singlephotons}, for threshold values of 6~mV and 12~mV
and different LED pulse amplitudes ranging from 840~mV to 950~mV. The diamond shaped markers represent the measured count rates of the CPM. Assuming the mentioned QE of $(3.0\pm0.3)$\% for the CPM, the photon rate incident onto the active area of the APD has been determined and is shown as the green band.\\
Comparing the count rates of the APD at these two thresholds (6~mV: blue circles, 12~mV: purple triangles) with the incident photon rates, a value of the photo detection efficiency can be extracted for each LED amplitude. The average of these values gives a photo detection efficiency of $(67\pm7)$\% for 6~mV threshold which fits well with the QE of 68\% at 628~nm interpolated from the datasheet of the S0223 APDs~\cite{RMD03}. At 12~mV threshold the photo detection efficiency drops to $(41\pm4)$\%.
\paragraph{Single photon detection efficiency at 1020~nm}\label{sec::1020nm}
\nl
For near infrared (NIR) wavelengths, a combination of a light bulb and a grating spectrograph is used as the light source. The characteristics of the grating spectrograph have been measured using a calibrated \emph{Thorlabs FDS 100-CAL} photodiode~\cite{Kus11}. The light from the spectrograph was subsequently attenuated by optical slits in the setup. Knowing the relative light intensities at different wavelengths one can calibrate the light source for single-photon emission in the NIR by determining the corresponding settings in the 
\begin{table}[h]
 \centering
  \caption[PDE and dark rates with spectrograph]{Results of the photo detection efficiency (PDE) determination for different wavelength with the grating spectrograph. The corresponding dark count rates are also shown.}
  \begin{tabular}{|c|c|c|c|c|}
   \hline
    & \multicolumn{2}{c|}{6~mV threshold} & \multicolumn{2}{c|}{12~mV threshold} \\
   \hline
   Wavelength  & APD PDE  & Average APD dark   & APD PDE  & Average APD dark \\
   (nm) & (\%) & rate (1/s) & (\%) & rate (1/s) \\
   \hline
    560 & $61 \pm 12$ &              & $37 \pm 7$  &              \\
    628 & $74 \pm 15$ & $\approx230$ & $45 \pm 9$  & $\approx0.7$ \\
   1020 & $13 \pm 3$  &              & $ 8 \pm 2$  &              \\
   \hline
  \end{tabular}
  \label{tab::NIRresults}
\end{table}
optical regime. The settings in the optical regime can again be checked with the CPM~1993~P channel photomultiplier. The additional uncertainty in the incident photon rates calculated at different wavelengths with this procedure is estimated to be on the 10\% level. APD count rates with and without light bulb were then taken at 560~nm, 628~nm and 1020~nm. The photo detection efficiency at 628~nm can be determined directly by using the previously calculated photon rate incident on the CPM. The incident photon rates at 560~nm and 1020~nm are calculated using the intensity ratios known from the above mentioned measurement with the calibrated photodiode.\\
The results regarding the photo detection efficiency and the dark count rate of the S0223 APD at 560~nm, 628~nm and 1020~nm are shown in table~\ref{tab::NIRresults} for the two threshold settings also used in the previous section. The values of the photo
detection efficiency at 628~nm agree well with the values obtained in the measurements with the LED as single-photon source. The photo detection efficiency of $(13\pm3)$\% measured for 1020~nm is lower than the QE of 20\% stated in the APD datasheet. \\
Especially when operating at low thresholds, we observe drifts in the dark count rates that are caused by fluctuations of the experimental parameters. These drifts have to be controlled by repeatedly measuring the dark count rates in between normal operation. Average values of the dark count rate at 6~mV and 12~mV threshold are given in table~\ref{tab::NIRresults}.\\
To estimate the measuring time per laser wavelength needed to detect a signal of the transition in hydrogen-like lead at the SPECTRAP experiment, one has to consider, that at the optimal threshold setting of 12~mV, as shown in figure~\ref{fig::oppoint}(c), the dark count rate is negligible compared to the dominating background from laser stray light in the SPECTRAP setup. Therefore
\begin{equation}
 D \rightarrow D \cdot \varepsilon \cdot \xi,
\end{equation}
where $\xi$ is the fraction of the light yield at the optical exits of the SPECTRAP setup that can be focused on a detector of a certain diameter. Equation~\ref{eqn::mintime} can then be rewritten to
\begin{equation}\label{eqn::time}
  t = \frac{n^2}{\varepsilon^2} \cdot \frac{\varepsilon \cdot D\cdot \xi}{S^2\cdot \xi^2}=\frac{n^2 \cdot D}{\varepsilon \cdot S^2 \cdot \xi}.
\end{equation}
In this case, with a $2\times2$~mm$^2$ size APD, $\xi=95$\%~\cite{Ham08}. The laser stray light background is estimated from a test measurement with Mg ions to be of the order $D=1.3\cdot10^{4}$~s$^{-1}$. Taking into account the expected fluorescence photon rate $S=6500$~s$^{-1}$ for the $^{207}\textrm{Pb}^{81+}$ hyperfine transition as listed in table~\ref{tab::photonrates}, a photodetection efficiency $\varepsilon=8$\% (see table~\ref{tab::NIRresults}) and a desired significance of $n=3$, equation~\ref{eqn::time} gives a measuring time $t\geq36$~ms per laser wavelength.
\section{Conclusion}\label{sec::conclusion}
We have shown that RMD S0223 APDs can be used as single-photon detectors at visible and near infrared wavelengths. 
At cryogenic temperatures below -160$^{\circ}$C relative gains of $M > 2\cdot 10^4$ have been achieved at a bias close to the breakdown voltage. The photondetection efficiency of the studied APDs at 628~nm wavelength is $(67\pm7)$\% and a dark count rate of about 230~$s^{-1}$ is observed. For near-infrared wavelengths around 1020~nm the detection efficiency is still about $(13\pm3)$\%. 
\section{Acknowledgments}
The author would like to thank the mechanics and electronics workshop of the Nuclear Physics Institute at the University of M\"unster for their support. This work has been supported by an R\&D contract with GSI and by BMBF under contract number 06MS9152I.

\begin{thebibliography}{99}
%
\bibitem{And10} Z. Andjelkovic et al.,
  \emph{Towards high precision in-trap laser spectroscopy of highly charged ions},
  \href{http://dx.doi.org/10.1007/s10751-009-0155-x}{Hyperfine Interactions {\bf 196}, 81-91 (2010)}
%
\bibitem{Sha01} V.M. Shabaev et al.,
  \emph{Towards a Test of QED in Investigations of the Hyperfine Splitting in Heavy Ions},
  \href{http://dx.doi.org/10.1103/PhysRevLett.86.3959}{Phys. Rev. Lett. {\bf 86}, 3959-3962 (2001)}
%
\bibitem{Qui01} W. Quint et al.,
  \emph{HITRAP: A Facility for Experiments with Trapped Highly Charged Ions},
  \href{http://dx.doi.org/10.1023/A:1011908332584}{Hyperfine Interactions {\bf 132}, 453-457 (2001)}
%
\bibitem{Vog05} M. Vogel et al.,
  \emph{Towards high precision in-trap laser spectroscopy of highly charged ions},
  \href{http://dx.doi.org/10.1063/1.2069742}{Review of Scientific Instruments {\bf 76}, 103102 (2005)}
%
\bibitem{RMD03} RMD, Inc.,
  \href{http://www.rmdinc.com/products/p006.pdf}{Silicon Avalanche Photodiode datasheet (2003)}
%
\bibitem{Mar00} J. Marler et al.,
  \emph{Studies of avalanche photodiode performance in a high magnetic field}
  \href{http://dx.doi.org/10.1016/S0168-9002(99)01382-0}{Nucl. Instr. and Meth. {\bf A 449}, 311-313 (2000)}
%
\bibitem{Gen11} T.R. Gentile et al.,
  \emph{Magnetic field effects on large area avalanche photodiodes at cryogenic temperatures}
  \href{http://dx.doi.org/10.1016/j.nima.2010.08.061}{Nucl. Instr. and Meth. {\bf A 652}, 520-523 (2011)} 
%
\bibitem{Ham08} D. Hampf,
  \emph{Untersuchung der APD S1315 von RMD im Hinblick auf ihren Einsatz als Detektor am SPECTRAP Experiment},
  \href{http://www.uni-muenster.de/Physik.KP/AGWeinheimer/Files/theses/Diplom_Daniel_Hampf.pdf}{Diploma thesis, Westf\"alische Wilhelms-Universit\"at M\"unster (2008)}
%
%
\bibitem{Wei92} C. Weinheimer et al.,
  \emph{Measurement of energy resolution and dead layer thickness of LN$_2$-cooled PIN photodiodes},
  \href{http://dx.doi.org/10.1016/0168-9002(92)90872-2}{Nucl. Instr. and Meth. A {\bf 311}, 273-279 (1992)};
  Diploma thesis M. Schrader, Johannes Gutenberg Universit\"at Mainz (1990)
%
\bibitem{Vis01} Vishay Siliconix,
  \href{http://www.datasheetcatalog.org/datasheet/vishay/70249.pdf}{U430 Matched N-Channel Pair datasheet (2001)}
%
\bibitem{Ber68} G. Bertolini and A. Coche,
  \emph{Semiconductor detectors}, North-Holland Pub. Co., Amsterdam (1968)
%
%
\bibitem{Per1} Perkin Elmer,
\href{http://www.excelitas.com/downloads/dts_pe_pd_02_channelphotomultipliersmodules.pdf}{CPM datasheet}
%
\bibitem{Sha09} P. Shagin et al.,
  \emph{Avalanche Photodiode for liquid xenon scintillation: Quantum Efficiency and gain}, 
  \href{http://iopscience.iop.org/1748-0221/4/01/P01005}{JINST 4 P01005 (2009)}
%
\bibitem{Kus11} F. Kuschewski,
  \emph{Untersuchung von Avalanche Photodioden zum Einzelphotonennachweis},
  \href{http://www.uni-muenster.de/Physik.KP/AGWeinheimer/Files/theses/Bachelor_Frederik_Kuschewski.pdf}{Bachelor thesis, Westf\"alische Wilhelms-Universit\"at M\"unster (2011)}
%
\end{thebibliography}
\end{document}